# On the energy efficiency of Laser-based Optical Wireless Networks

Walter Zibusiso Ncube, Ahmad Adnan Qidan, Taisir El-Gorashi and Jaafar M. H. Elmirghani
School of Electronic and Electrical Engineering University of Leeds, Leeds, United Kingdom
{el18wzn, A.A. Qidan, t.e.h.elgorashi, J.M.H.Elmirghani}@leeds.ac.uk

*Abstract*—Optical wireless Communication (OWC) is a strong candidate in the next generation (6G) of cellular networks. In this paper, a laser-based optical wireless network is deployed in an indoor environment using Vertical Cavity Surface Emitting Lasers (VCSELS) as transmitters serving multiple users. Specifically, a commercially available low-cost VCSEL operating at 850nm wavelength is used. Considering the confined coverage area of each VCSEL, an array of VCSELs is designed to transmit data to multiple users through narrow beams taking into account eye safety regulations. To manage multi-user interference (MUI), Zero Forcing (ZF) is implemented to maximize the multiplexing gain of the network. The energy efficiency of the network is studied under different laser beam waists to find the effective laser beam size that results in throughput enhancement. The results show that the energy efficiency increases with the laser beam waist. Moreover, using micro lenses placed in front of the VCSELs leads to significant increase in the energy efficiency.

*Keywords*—- *Optical Wireless Networks, Vertical Cavity Surface Emitting Lasers, Zero Forcing, Interference Management, Energy Efficiency*

## I. INTRODUCTION

Global internet traffic grows exponentially due to the massive use of internet-based applications on daily basis. Given that, current radio frequency (RF) networks may fail to support unprecedented user demands [1], [2]. Optical wireless Communication (OWC) presents a key technology to complement RF wireless communication, providing high aggregate data rates in a range of Gigabit per second [3]–[9]. OWC uses the wide, free, unregulated visible and infrared light spectrums. Conventional OWC systems consider the use of Light Emitting Diodes (LEDs) or Laser Diodes (LDs) to provide both functions: illumination and communication [10], [10].Such systems usually use low-cost Solid-State Devices (SSD) that have the dual nature of illumination and communication [11]–[13]. The performance of LED-based OWC systems is limited due to the low modulation speed of the LED [14]. Therefore, alternative light sources are needed to unlock high communication speeds.

Laser Diodes are characterized by their high modulation speed, and hence they have the potential to exploit the massive licence-free bandwidth of the optical spectrum and provide high data rates [15]. Compared to other lasers, vertical-cavity surface-emitting lasers (VCSELs) are gaining much traction as light sources for short-distance communication due to several advantages including low-cost and high-bandwidth [16], [17], [18]. Moreover, VCSELs are also considered as energy efficient transmitters [5], [19], [20]. In [21], it is shown that VCSELs have energy efficiency of 140 mW/Tb/s whilst operating at 1060 nm and achieving 10 Gb/s. To elaborate further, active materials with better gain properties and VCSELs with longer wavelengths using less energy per generated photon can improve the performance of OWC systems considerably [21]. In the context of energy efficiency, VCSELs operating at a wavelength range of 850 nm up to 1550 nm have been widely used to transmit data at high communication speeds. It motivates this work to use an array of VCSELs for transmitting data to multiple users whilst adhering to human eye safety requirements.

In this paper, a laser-based optical wireless network is considered to provide high energy efficiency. It is worth mentioning that the energy efficiency of a system can be expressed by the Consumption Factor (CF), in bits-per-Joule, which is defined as the ratio of the transmitted data over the consumed energy [22]. A higher CF means less energy is consumed to transmit the data. This term was used as a standard base for the design and analysis of communication systems in [23], [24], [25], [26]. We first define our system model in which multiple arrays are deployed on the ceiling to serve multiple users. In this sense, the application of Zero Forcing (ZF) is considered to manage multi-user interference. Then, mathematical equations are derived to determine the beam profile of the laser. Finally, we use micro lenses to maximize the received power of the user. The results show that an array of VCSELs can achieve high energy efficiency, which increases with the beam waist of the laser. Moreover, the use of micro lenses can further enhance the energy efficiency of the network.

## II. SYSTEM MODEL

We consider a downlink OWC system in an indoor environment. On the ceiling, multiple optical access points (APs), $A$, are deployed to provide service for multiple users distributed on the receiving plane. Each optical AP comprises multiple VCSELs arranged in an $N \times N$ array with micro lenses placed in front of the VCSELs. In this work, multiple users, $U$, are randomly distributed over the communication floor, which is 1 m above the ground as shown in Figure 1. Moreover, each user is equipped with a wide Field of View (FOV) optical detector that points towards the ceiling. Our aim in this study is to determine the energy efficiency of VCSEL transmitters in OWC systems. It is worth pointing out that a controller, which can be in the optical line terminal in the room, is assumed to have knowledge of user-distribution, in addition to the information required to manage the resources of the network. This controller has the ability to respond to the slow movement of the users in the indoor environment, and it can update the resource allocation accordingly. In Table 1, the room dimensions as well as the parameters of the transmitter and receiver are given. Note that, conventional On-Off Keying (OOK) modulation is assumed, and only Line-Of-Sight (LOS) [27], components are taken into consideration due to the fact that they represent the largest portion of the power received.

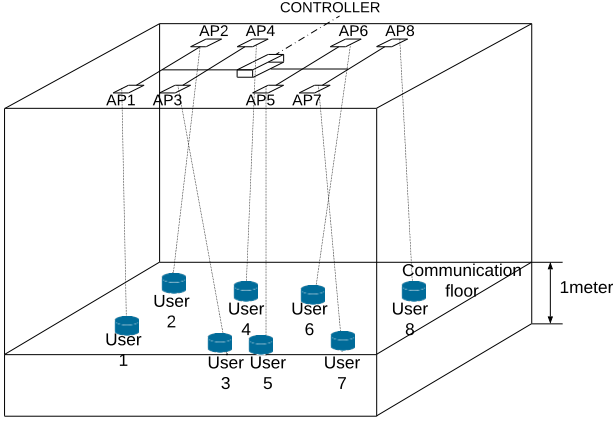

Figure 1: System Model

Managing multi-user interference (MUI) is a crucial issue in this system that must be addressed efficiently to maximize the overall sum rates of the network, achieving high energy efficiency. Therefore, a well-known transmit precoding scheme referred to as ZF is implemented to handle MUI and guarantee that each user can decode its information correctly. Basically, MUI can be managed at the transmitter end of the communication system using precoding methods [28]. In [28]–[39] various precoding methods have been derived to enhance the spectral efficiency of RF networks. These methods must go through certain developments prior to their application in OWC, which involves high complexity due to the unique nature of the optical signal. Using ZF, the precoding algorithm tries to remove MUI at a given user by creating a pre-coder, $G_i$, that is orthogonal to the channel matrices of other interfering users. This process is introduced below in detail as in [28], [39].

Given an aggregate channel matrix $H_n^{a,}$, we construct a ZF precoder $G_i$ that is orthogonal to $H_n^{a,}$, leading to:

$$G_i^{a,} \times H_n^{a,} = 0, \forall n \neq i \quad (1)$$

The ZF constraint implies that:

$$G_i^{a,} H_n^{a,} = \begin{vmatrix} \sqrt{q_1} & & \\ & \ddots & \\ & & \sqrt{q_n} \end{vmatrix} = diag\{\sqrt{q}\} \quad (2)$$

where $\sqrt{q} = [\sqrt{q_1} \sqrt{q_2} \cdots \sqrt{q_u}]$ whose elements represents the channel gain of the users.
Considering that $H^-$ denotes the generalised inverse of H, which can be any matrix that satisfies $(H^+H)^-H^+ = H^-$. It should be noted that G is both a pseudo inverse and a non-unique left inverse. Note that the implementation of ZF is relatively straightforward with low complexity. However, its performance is subject to several limitations including noise enhancement and the need for channel state information (CSI) at transmitters [40], which represents a challenge in wireless communication since it requires the information and data of the users to be exchanged among all the APs in the network.

## III. VERTICAL CAVITY SURFACE EMITTING LASERS (VCSEL))

VCSELs are semiconductor-based laser diodes that emit light from their surfaces in a direction normal to the substrate. These IR lasers were originally developed by Kenichi Iga and they can be used in imaging, sensing, and scanning in fields such as machine vision and medicine [41]–[43]. Examples of these applications are facial recognition, 3D sensing and 2D imaging. VCSELs have relatively low production cost, and multiple VCSELs can be processed on a single wafer, and they can be tested at various stages in wafer form [44]. VCSELs also offer favourable characteristics such as longitudinal single-mode operation, large modulation bandwidth and high-quality circular beams. Furthermore, two-dimensional VCSEL arrays having hundreds of individual light sources can be created to increase the output power and long-term reliability, while expanding the coverage area. Therefore, they can be considered for use as transmitters in OWC.

The output beam profile of multi-mode VCSELs can be expressed as the superposition of Laguerre-Gaussian mode as presented in [45], and the radial power for each mode is given by

$$|U_{p,l}(r,z)|^2 = (A_p^l)^2 \times \frac{w_0^2}{w(z)^2} \times (\frac{w_0^2}{w(z)^2})^l \times (L_p^l(2\frac{r^2}{w(z)^2}))^2 \times exp(-\frac{2r^2}{w(z)^2}) \quad (3)$$

where $w_0$ is the beam waist at z = 0 and $w_z$ is the beam waist at distance z. Moreover, $L_p^l$ is the generalised Laguerre polynomials, which can be given by

$$L_p^l(x) = \sum_{m=0}^{p}(-1)^m \times \frac{(p+l)!}{(p-m)!(l+m)!m!} \times x^m \quad (4)$$

where $A_p^l$ is the normalisation constant used to ensure that the total mode power of each mode is 1, i.e.,

$$\int_0^\infty \int_0^{2\pi} |U_{p,l}(r,\varphi,z)|^2 r dr d\varphi \quad (5)$$

$$A_p^l = \frac{1}{w_0} \times \sqrt{\frac{2p!}{\pi(p+l)!}} \quad (6)$$

The integers $p = 0, 1$ and $l = 0, 1, 2, 3$ are radial and azimuthal indices, respectively. The radius of the beam spot is calculated as:

$$w_z = w_0 \sqrt{1 + (\frac{z}{z_r})^2} \quad (7)$$

where the variable $z_r$ is the Raleigh range, it is calculated as:

$$z_r = \frac{\pi w_0^2}{\lambda} \quad (8)$$

It should be noted that for large values of $z$, $z \gg z_r$, $w_z$ approaches asymptotic value such that:

$$w_z \approx \frac{z w_0}{z_r} \quad (9)$$

This reduces to:
$$w_z \approx \frac{z\lambda}{w_0 \pi} \quad (10)$$

The radius of the phase front is planar initially but expands in the direction of propagation. At a distance $z$, the radius of the phase front is given by:

$$R_z = z\left(1 + \left(\frac{\pi w_0^2}{\lambda}\right)^2\right) \quad (11)$$

The divergence half angle of the beam is given by:

$$\theta = \tan^{-1}\frac{w_z}{z}. \quad (12)$$

A micro lens array is employed in this study. Note that, the use of this lens array is proposed taking into account eye safety calculations, it modifies the beam profile of the VCSEL. Using the thin lens approximation, the micro lens results in the VCSEL output beam being transformed in accordance to the ABCD formulation [45], [46].

Consequently, the beam waist becomes minimum at distance $d_2$. This is calculated as:

$$d_2 = \frac{\frac{1}{f} - \left(1 - \frac{d_1}{f}\right)d_1\left(\frac{\lambda}{\pi w_0^2}\right)^2}{\frac{1}{f^2} + \left(1 - \frac{d_1}{f}\right)^2\left(\frac{\lambda}{\pi w_0^2}\right)^2} \quad (13)$$

The minimum waist is calculated as:
$$w_l = \frac{\lambda f}{\pi w_0 \sqrt{1 + \left(1 - \frac{d_1}{f}\right)^2 \frac{\lambda^2}{\pi^2 w_0^4} f^2}}. \quad (14)$$

The beam divergence after the lens is given by:
$$\theta_2 = \frac{\theta}{k} \quad (15)$$

where $k$ is the magnification factor of the lens, i.e.,
$$k = \frac{w_2}{w_0} \quad (16)$$

## IV. Considerations of VCSEL Array Design

In this work, each optical AP is composed of an $N \times N$ VCSEL array. In this context, the total transmitted power is given by the sum of the power from all the individual VCSEL emitters. It is worth mentioning that the heat resulting from the emissions of each VCSEL must be considered in the design of such an array. Given that, the maximum radiated power of each VCSEL is restricted by the cooling system due to the significant thermal power density of the VCSEL array. This can be controlled by adjusting the pitch distance between the VCSELs of the array. In other words, the heat can be decreased by increasing the pitch distance. However, increasing the pitch distance may negatively impact the beam quality and the speed of transmission (however cooling is beyond the scope of this paper).

To guarantee that the operation of the VCSELs array satisfies the eye safety requirements, each VCSEL unit must be considered when determining the Maximum Permissible Emission (MPE) for eye safety. For the VCSEL source to be safe for human eyes, a fraction of the power, $\eta$, from the Gaussian beam received by the eye at the most hazardous position must be less than the product of the maximum permissible exposure and the area of the eye aperture. In addition, the transmitted power is subject to the constraint $P_u^{a,} \leq P_{max}^{'}$, where $\forall u, \in U, \forall a, \in A$ and $P_{max}^{'}$ is the maximum transmit power of the VCSEL permitted in the network due to eye safety and heating constraints. According to [9], the maximum transmit power from the VCSEL source can be determined as below.

The distance for which 86% of the power passes into the pupil is calculated as:

$$d_{86\%} = \frac{\pi w_o}{\lambda}\sqrt{\frac{-2r_p^2}{\ln(1-0.86)}} \quad (17)$$

If $d_{86\%}$ is less than 0.1m then MHP is set to 0.1m. The subtense angle is given by:

$$\alpha = 2\tan^{-1}\left(\frac{w_o}{MHP}\right). \quad (18)$$

Then, the maximum power for eye safety is calculated as:
$$P_{max} = \frac{1}{\eta} MPE(\pi r_p^2) \quad (19)$$

where
$$\eta = 1 - \exp\left(\frac{-2r_p^2(\pi w_0^2)^2}{(\lambda \times MHP)^2}\right) \quad (20)$$

For a VCSEL source with a lens, $d_2$ and $w_l$ are used as the new beam waist and beam divergence (see (13) and (14)).

## V. Channel model

In this work, the channel is simulated using a ray tracing algorithm. The simulation is carried out using MATLAB. All the parameters of the simulation are listed in Table 1. The total received power $P_r$ at the receiver considering the LOS component ($PLOS$), first-order reflections $P_{FST}$ and second-order reflections $P_{SEC}$ [47] ,can be expressed as

$$P_r = \sum_{i=1}^{S} P_{LOS} + \sum_{i=1}^{M} P_{FST} + \sum_{i=1}^{F} P_{SEC}. \quad (21)$$

For the purposes of this study, only LOS links are considered, for the sake of simplicity. The system performance is determined by the SINR. Moreover, three noise sources are considered in the determination of the noise penalties. That is, the SNR is calculated as:

$$\gamma_u^{a,} = \frac{signal}{noise} = \frac{P_{r,u}^{a,} R}{\delta_t}. \quad (22)$$

The electrical signal power received by user $u$ from access point $a$ is calculated as:

$$P_{r,u}^{a,} = RP_u^{a,} H_u^{a,} G_i^{a,} \ (i \neq u) \quad (23)$$

Table 1: System Parameters

| Parameters | Configurations |
|---|---|
| VCSEL | |
| Beam waist, $W_0$ | 5 $\mu m$ |
| Pitch Distance | 10 $\mu m$ |
| Link Distance | 2 m |
| VCSELs per transmitter | 25 |
| Lens | Yes |
| Lens focal length | 0.127 mm |
| VCSEL to lens | 0.133 mm |
| VCSEL wavelength | 850 nm |
| VCSEL Bandwidth | 5 GHz |
| Laser noise PSD | -155 dB/Hz |
| Lens refractive index | 1.5 |
| Load resistance | 50 Ohms |
| TIA noise figure | 5 dB |
| FEC limit | $10^{-3}$ |
| Semi-angle of reflection element at half power | 15° |
| Room | |
| Width × Length × Height | 5m × 5m × 3m |
| Number of optical transmitter units | 4 |
| Optical transmitter locations (x, y, z) | (3m,3m,3m),(1m,3m,3m), (3m,1m,3m),(1m,1m,3m), (1m,5m,3m),(1m,7m,3m), (3m,5m,3m),(3m,7m,3m) |
| Responsivity | 0.4 A/W |
| Area of the photodetector | 2 $cm^2$ |
| Receiver noise current spectral density | 4.47 pA/√Hz |
| Bias current | 9 mA |
| Peak-to-peak driving amplitude voltage | 0.9 V |
| Receiver bandwidth | 1.75 GHz |

The total noise penalty is calculated as in [48]:

$$\delta_T = \sqrt{\left(\delta_{Rx}^2 + \delta_{Th}^2 + \sigma rin_n^{2\,a} + \delta_u^{b\,2}\right)} \quad (24)$$

where:
$$\delta_u^{b\,2} = 2eB_oB_e(RP_u^b h_u^b) \quad (25)$$
$$\delta_{Th}^2 = 4K_BTN_f/R_l \quad (26)$$
$$\sigma rin_n^{2\,a} = BN_{rin}(RPt_u^a h_u^a)^2 \quad (27)$$
$$\delta_{Rx}^2 = N_{pr}B_e \quad (28)$$

$e$: is the electron charge.

$B_o$ and $B_e$: are the optical and electrical bandwidth.
$K_B$: is the Boltzmann constant.
$N_f$: is the noise figure of the amplifier.
$T$: is the temperature.
$R_L$: is the load resistance.

$$N_{rin} = \sqrt{10^{\frac{RIN}{10}}B}$$

RIN: is the RIN value.
$N_{pr}$: is the preamplifier noise power density in ($A^2$/Hz).

## VI. RESULTS

In this section the performance of the network is determined in terms of the sum rate and energy efficiency assuming various beam waist values, which can highly affect the received power of the user.

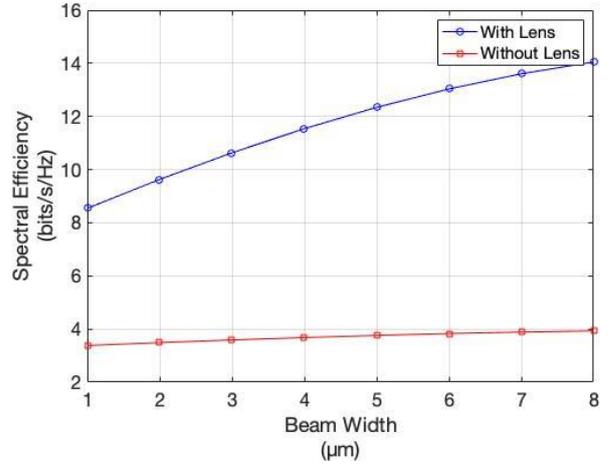

Figure 2: the spectral efficiency of the network versus a range of laser beam waist.

In Fig. 2. the sum rate of the network is depicted against a range of beam waist values from 1 $\mu m$ to 8 $\mu m$, with and without the use of the lens. It is shown that the sum rate of the network increases with the beam waist regardless of the scenario considered due to the fact that increasing the beam waist results in focusing the radiated power towards the users, and therefore the SNR of the user increases considerably. It is worth pointing out that selecting a high beam waist might violate the eye safety requirements. Therefore, the trade-off between the increase in the SNR and eye safety must be considered prior to adjusting the laser beam waist. The figure further shows that the use of the lens enhances the overall sum rate of the network significantly where each user can receive a larger portion of the radiated power compared to the scenario of serving users without placing lenses in front of the VCSEL units.

Fig. 3 shows the energy efficiency of the network versus the laser beam waist considering two different scenarios. It can be seen that the use of the lens results in high energy efficiency compared to the other scenario without the use of the lens. This is because the radiated power is focused towards the optical detectors of the users, which leads to SNR enhancement. Despite the complex design of the transmitters with micro lenses, it is worth implementing such an approach where the energy efficiency can be enhanced without consuming more input power. Moreover, a larger beam waist increases the energy efficiency of the network in both scenarios due to the increase in the user's received power.

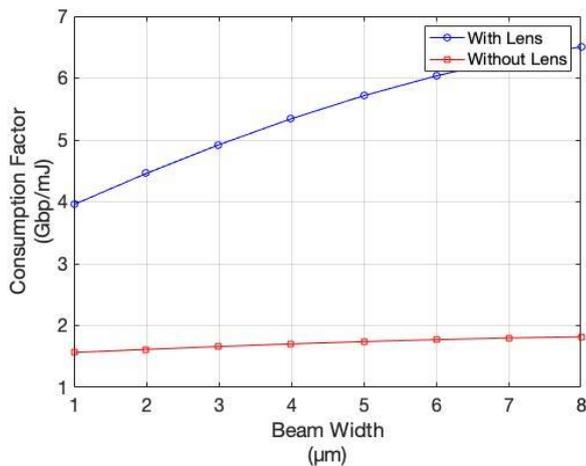

Figure 2: the energy efficiency of the network versus a range of laser beam waist.

VII. CONCLUSIONS

In this paper the performance of a laser-based optical wireless network is evaluated in terms of energy efficiency considering the use of micro lenses. We first define the system model, which is composed of multiple optical APs serving multiple users. To manage MUI, ZF is implemented to support multiple access services where each user can decode its information while cancelling the interference received due to the transmission to other users, which results in maximizing the sum rate of the network. Then, we derive mathematical expressions for the laser beam profile to calculate the radiated power of the laser with and without micro lenses. The results show that the use of the lens increases the sum rate and energy efficiency of the network due to increasing the SNR of the user. Moreover, the beam waist of the laser can affect the performance of the network where a high value of the beam waist results in a high received power. For future work, the formulation of various power allocation optimization problems might be considered to maximize the overall SNR and sum rates of laser-based optical wireless networks, while satisfying the eye safety requirements.

ACKNOWLEDGMENTS

This work has been supported by the Engineering and Physical Sciences Research Council (EPSRC), in part by the INTERNET project under Grant EP/H040536/1, and in part by the STAR project under Grant EP/K016873/1 and in part by the TOWS project under Grant EP/S016570/1. All data are provided in full in the results section of this paper.

REFERENCES

[1] A. A. Qidan, M. Morales-Céspedes, T. El-Gorashi, and J. M. H. Elmirghani, "Resource Allocation in Laser-based Optical Wireless Cellular Networks," *GLOBECOM 2021 - IEEE Glob. Commun. Conf.*, 2021.

[2] S. H. Younus, A. A. Al-Hameed, A. T. Hussein, M. T. Alresheedi, and J. M. H. Elmirghani, "Parallel Data Transmission in Indoor Visible Light Communication Systems," *IEEE Access*, vol. 7, pp. 1126–1138, 2019.

[3] M. T. Alresheedi and J. M. H. Elmirghani, "10 Gb/s Indoor Optical Wireless Systems Employing Beam Delay, Power, and Angle Adaptation Methods With Imaging Detection," *J. Light. Technol.*, vol. 30, no. 12, pp. 1843–1856, Jun. 2012.

[4] A. T. Hussein and J. M. H. Elmirghani, "10 Gbps mobile visible light communication system employing angle diversity, imaging receivers, and relay nodes," *J. Opt. Commun. Netw.*, vol. 7, no. 8, pp. 718–735, 2015.

[5] P. Westbergh, E. P. Haglund, E. Haglund, R. Safaisini, J. S. Gustavsson, and A. Larsson, "High-speed 850 nm VCSELs operating error free up to 57 Gbit/s," *Electron. Lett.*, vol. 49, no. 16, pp. 1021–1023, 2013.

[6] M. T. Alresheedi and J. M. H. Elmirghani, "High-Speed Indoor Optical Wireless Links Employing Fast Angle and Power Adaptive Computer-Generated Holograms With Imaging Receivers," *IEEE Trans. Commun.*, vol. 64, no. 4, pp. 1699–1710, 2016.

[7] F. E. Alsaadi and J. M. H. Elmirghani, "High-Speed Spot Diffusing Mobile Optical Wireless System Employing Beam Angle and Power Adaptation and Imaging Receivers," *J. Light. Technol.*, vol. 28, no. 16, pp. 2191–2206, 2010.

[8] M. T. Alresheedi and J. M. H. Elmirghani, "Performance Evaluation of 5 Gbit/s and 10 Gbit/s Mobile Optical Wireless Systems Employing Beam Angle and Power Adaptation with Diversity Receivers," *IEEE J. Sel. Areas Commun.*, vol. 29, no. 6, pp. 1328–1340, 2011.

[9] A. A. Qidan, M. Morales-Céspedes, A. G. Armada and J. M. H. Elmirghani, "Resource Allocation in User-Centric Optical Wireless Cellular Networks Based on Blind Interference Alignment," in Journal of Lightwave Technology, vol. 39, no. 21, pp. 6695-6711, 1 Nov.1, 2021, doi: 10.1109/JLT.2021.3107140.

[10] S. H. Younus, A. T. Hussein, M. Thamer Alresheedi, and J. M. H. Elmirghani, "CGH for Indoor Visible Light Communication System," IEEE Access, vol. 5, pp. 24988–25004, 2017.

[11] O. Z. Alsulami et al., "Optimum resource allocation in 6G optical wireless communication systems," arXiv, 2020.

[12] H. Haas, J. Elmirghani, I. White, and H. Haas, "Optical wireless communication," Philos. Trans. R. Soc. A, vol. 378, no. 2169, pp. 1–11, 2020.

[13] A. A. Al-Hameed, S. H. Younus, A. T. Hussein, M. T. Alresheed, and J. M. H. Elmirghani, "LiDAL: Light Detection and Localization," IEEE Access, vol. 7, pp. 85645–85687, 2019.

[14] A. S. Elgamal, O. Z. Aletri, A. A. Qidan, T. E. H. El-Gorashi and J. M. H. Elmirghani, "Reinforcement Learning for Resource Allocation in Steerable Laser-Based Optical Wireless Systems," 2021 IEEE Canadian Conference on Electrical and Computer Engineering (CCECE), 2021, pp. 1-6, doi: 10.1109/CCECE53047.2021.9569123.

[15] A. T. Hussein and J. M. H. Elmirghani, "Mobile Multi-Gigabit Visible Light Communication System in Realistic Indoor Environment," J. Light. Technol., vol. 33, no. 15, pp. 3293–3307, 2015.

[16] A. A. Qidan, T. El-Gorashi1, and J. M. H. Elmirghani, "Artificial Neural Network for Resource Allocation in Laser-based Optical wireless Networks," *ICC 2022 - IEEE International Conference on Communications*, 2021, pp. 1-7.

[17] C.-H. Yeh, L.-Y. Wei, and C.-W. Chow, "Using a Single VCSEL Source Employing OFDM Downstream Signal and Remodulated OOK Upstream Signal for Bi-directional Visible Light Communications," Sci. Rep., vol. 7, no. 1, p. 15846, 2017.

[18] R. Stevens, "Modulation Properties of Vertical Cavity Light Emitters Doctoral Thesis by," Technology, no. November, 2001.

[19] J. M. Wun, J. W. Shi, J. C. Yan, J. J. Chen, and Y. J. Yang, "Oxide-Relief and Zn-Diffusion 850 nm Vertical-Cavity Surface-Emitting Lasers with Extremely Small Power Consumption and Large Bit

Rate-Distance Product for 40 Gbit/sec Operations," Opt. Fiber Commun. Conf. Fiber Opt. Eng. Conf. 2013 (2013), Pap. OM3K.6, p. OM3K.6, Mar. 2013.

[20] D. M. Kuchta et al., "A 71-Gb/s NRZ modulated 850-nm VCSEL-based optical link," IEEE Photonics Technol. Lett., vol. 27, no. 6, pp. 577–580, 2015.

[21] P. Moser, Energy-Efficient VCSELs for Optical Interconnects. Springer International Publishing, 2016.

[22] J. N. Murdock and T. S. Rappaport, "Consumption factor and power-efficiency factor: A theory for evaluating the energy efficiency of cascaded communication systems," IEEE J. Sel. Areas Commun., vol. 32, no. 2, pp. 221–236, 2014.

[23] H. M. Kwon and T. G. Birdsall, "Channel Capacity in Bits Per Joule," IEEE J. Ocean. Eng., vol. 11, no. 1, pp. 97–99, 1986.

[24] L. Teixeira, F. Loose, C. H. Barriquello, V. A. Reguera, M. A. D. Costa, and J. M. Alonso, "On energy efficiency of visible light communication systems," IEEE J. Emerg. Sel. Top. Power Electron., vol. 9, no. 5, pp. 6396–6407, 2021.

[25] S. Singh, N. Saxena, A. Roy, and H. S. Kim, "Energy Efficiency in Wireless Networks–a Composite Review," IETE Tech. Rev. (Institution Electron. Telecommun. Eng. India), vol. 32, no. 2, pp. 84–93, 2015.

[26] N. D. Chatzidiamantis, L. Georgiadis, H. G. Sandalidis, and G. K. Karagiannidis, "An Efficient Power Constrained Transmission Scheme for Hybrid OW/RF Systems."

[27] F. E. Alsaadi and J. M. H. Elmirghani, "Beam power and angle adaptation in multibeam 2.5 Gbit/s spot diffusing mobile optical wireless system," IEEE J. Sel. Areas Commun., vol. 28, no. 6, pp. 913–927, 2010.

[28] T. V. Pham, H. Le-Minh, and A. T. Pham, "Multi-User Visible Light Communication Broadcast Channels With Zero-Forcing Precoding," Elsevier, vol. 65, pp. 25–35, 2018.

[29] M. Joham, W. Utschick, and J. A. Nossek, "Linear Transmit Processing in MIMO Communications Systems," IEEE Trans. Signal Process., vol. 53, no. 8, pp. 2700–2712, 2005.

[30] H. Shen, Y. Deng, W. Xu, and C. Zhao, "Rate-maximized zero-forcing beamforming for VLC multiuser MISO downlinks," IEEE Photonics J., vol. 8, no. 1, pp. 1–13, 2016.

[31] A. Agarwal and S. K. Mohammed, "Achievable Rate Region of the Zero-Forcing Precoder in a MU-MISO Broadcast VLC Channel with Per-LED Peak Power Constraint and Dimming Control," J. Light. Technol., vol. 35, no. 19, pp. 4168–4194, 2017.

[32] L. Song and X. Wang, "Beam selection methods for massive MIMO systems with hybrid radio frequency and baseband precoding," 2015 IEEE 82nd Veh. Technol. Conf. VTC Fall 2015 - Proc., vol. 24, no. 3, pp. 528–541, 2016.

[33] D. Gesbert, M. Kountouris, R. W. Heath, C. B. Chae, and T. Sälzer, "Shifting the MIMO Paradigm," IEEE Signal Process. Mag., vol. 24, no. 5, pp. 36–46, 2007.

[34] D. Bartolomé and A. I. Pérez-Neira, "Spatial scheduling in multiuser wireless systems: From power allocation to admission control," IEEE Trans. Wirel. Commun., vol. 5, no. 8, pp. 2082–2091, 2006.

[35] G. Del Galdo and M. Haardt, "Comparison of zero-forcing methods for downlink spatial multiplexing in realistic multi-user MIMO channels," IEEE Veh. Technol. Conf., vol. 59, no. 1, pp. 299–303, 2004.

[36] C. Sun and J. Ge, "Low complexity suboptimal user selection algorithm for multiuser MIMO systems with block diagonalization," Wirel. Pers. Commun., vol. 75, no. 4, pp. 1937–1946, 2014.

[37] J. Wang, Z. Liu, Y. Wang, and X. You, "Performance of the Zero Forcing precoding MIMO broadcast systems with channel estimation errors," J. Electron., vol. 24, no. 4, pp. 490–495, 2007.

[38] H. Sifaou, K. H. Park, A. Kammoun, and M. S. Alouini, "Optimal linear precoding for indoor visible light communication system," arXiv, 2017.

[39] A. Wiesel, Y. C. Eldar, and S. Shamai, "Zero-forcing precoding and generalized inverses," IEEE Trans. Signal Process., vol. 56, no. 9, pp. 4409–4418, 2008.

[40] Qidan, A.; El-Gorashi, T. and Elmirghani, J. (2022). Towards Terabit LiFi Networking. In Proceedings of the 10th International Conference on Photonics, Optics and Laser Technology, ISBN 978-989-758-554-8, ISSN 2184-4364, pages 203-212. DOI: 10.5220/0010955000003121

[41] A. Adibi et al., VCSELs Fundamentals, Technology and Applications of Vertical-Cavity Surface-Emitting Lasers, vol. 34, no. 8. 2013.

[42] K. Hsuan-yun et al., "VCSELs for optical OFDM transmission," vol. 25, no. 14, pp. 16347–16363, 2017.

[43] W. H. Hofmann, P. Moser, and D. Bimberg, "Energy-efficient VCSELs for interconnects," IEEE Photonics J., vol. 4, no. 2, pp. 652–656, 2012.

[44] J. Lavrencik, "Vcsel-Based Multimode Fiber Optical Links For High Capacity Interconnects," 2020.

[45] N. Bamiedakis, "Note on VCSEL output beam and eye safety calculations," Univ. Cambridge, pp. 1–12, 2020.

[46] A. M. Rashed, "Source misalignment in multimode polymer tapered waveguides for optical backplanes," Opt. Eng., vol. 46, no. 1, p. 015401, 2007.

[47] F. E. Alsaadi, M. Nikkar, and J. M. H. Elmirghani, "Adaptive mobile optical wireless systems employing a beam clustering method, diversity detection, and relay nodes," IEEE Trans. Commun., vol. 58, no. 3, pp. 869–879, 2010.

[48] H. Kazemi, E. Sarbazi, M. D. Soltani, M. Safari, and H. Haas, "A Tb/s indoor optical wireless backhaul system using VCSEL arrays," IEEE Int. Symp. Pers. Indoor Mob. Radio Commun. PIMRC, vol. 2020-Augus, pp. 1–30, 2020.